\documentclass[11pt]{article}

\usepackage[margin=1.1in]{geometry}
\usepackage[T1]{fontenc}
\usepackage[utf8]{inputenc}
\usepackage{amsmath,amssymb}
\usepackage{graphicx}
\usepackage{booktabs}
\usepackage{microtype}
\usepackage{float}
\usepackage[hidelinks]{hyperref}
\usepackage{cite}
\usepackage{tikz}
\usetikzlibrary{positioning,fit,arrows.meta,calc}

\title{Reflector: Arrangement-Aware Harmonic Retrieval\\for Sample-Based Composition}

\author{Austin Rockman\\
  Independent Researcher\\
  \texttt{research@austinrockman.com}}

\date{}

\begin{document}

\maketitle

\begin{abstract}
Sample retrieval tools can help composers find harmonically compatible material, but querying from a fixed reference sample becomes less informative as arrangements evolve and the harmonic context shifts with each musical decision. We present Reflector, an interactive audio workstation that tracks harmonic combinations as they accumulate on the composer's timeline and adapts retrieval as the arrangement develops. The system is organized around a fixed interval-class oracle: a hand-designed table of weights that scores how pitch-class content combines between sources. An encoder trained entirely on synthetic audio learns to approximate the oracle in a 128-dimensional embedding space, where dot products stand in for compatibility scores at interactive speed. As the composer arranges material on a multi-track timeline, a sweep-line analysis discovers co-sounding regions, computes oracle-weighted centroids, and retrieves against the composite harmonic identity of the session as it evolves. Session centroids projected into a navigable 3-D space reveal structural harmonic relations across the composer's body of work. This paper is a systems account: we give the design rationale for each architectural decision, characterize Reflector's behavior through intrinsic measurements on a working sample library, and describe the implementation. The characterization yields a central finding: the learned embedding preserves the kernel's pairwise judgments while covering the whole library, something the kernel cannot do when used directly as a retrieval rule, because the embedding's normalized geometry cannot express the degenerate solutions that direct scoring favors. The entire pipeline runs locally with no copyrighted training data. Reflector is free, and the training pipeline is open source.
\end{abstract}

\section{Introduction and Related Work}\label{sec:introduction}

Music Information Retrieval (MIR) has produced effective tools for retrieving and navigating material along musical axes such as timbre, rhythm, harmony, and structure. Neural models for loop compatibility have learned to score whether two loops fit together, outperforming rule-based alternatives~\cite{Chen2020}. Context-conditioned retrieval has been explored for production tasks such as drum-sample suggestion, where a rendered mixdown of an incomplete song serves as the query~\cite{Lattner2022}. Psychoacoustic consonance models have been used for harmonic mixing and transposition search between pairs of audio signals~\cite{Gebhardt2016}. Context-to-stem compatibility has been studied with learned representations that retrieve a missing stem from a multi-instrumental context~\cite{Riou2024}. Products such as Sononym~\cite{Sononym} let composers navigate sample libraries by sonic similarity; XLN XO~\cite{XLNXO} pairs similar visualization with an integrated drum sequencer. Each of these systems extracts meaningful structure from its query, whether a single loop, a rendered mix, or a multi-track context.

Commercial platforms increasingly treat retrieval as compositional. Splice's Create~\cite{SpliceCreate} suggests complementary sounds in real time, adjusting as the composer modifies an arrangement; Output's Co-Producer~\cite{OutputCoProd} analyzes a session and retrieves fitting samples. Both operate over the platform's sample catalog, even where the composer's own audio seeds the search, and their methodologies are proprietary and unpublished. One purpose of this paper is to provide a public account of how an arrangement-aware retrieval system might be designed, trained, and operated end to end on a single machine.

A separate body of work measures tonal content itself. Harmonic pitch-class profiles~\cite{Gomez2006} and chroma features summarize pitch-class energy; tonal centroid and Tonnetz distances~\cite{Harte2006} and Tonal Interval Vectors~\cite{Bernardes2016} embed that energy in spaces where geometric distance reflects tonal relatedness, and transposition-invariant chroma matching~\cite{Muller2007} aligns tonal content across keys. Reflector builds on the same pitch-class substrate, and its front end is a constant-Q transform (CQT)~\cite{Brown1991,BrownPuckette1992} chroma pipeline.

In practice, the early stage of composition is an ideation process of browsing, testing, and arranging candidate materials, during which harmonic relevance shifts as new sounds are placed in relation to one another. Systems that track the evolving arrangement turn retrieval from a static lookup into a continuous participant in that process. To our knowledge, no published work has modeled a multi-track arrangement as an evolving set of co-sounding regions for harmonic retrieval, or examined how such representations persist across sessions as a record of compositional practice.

Reflector explores one axis of sample retrieval: harmonic compatibility over pitch-class content. A composer begins from a single sample, retrieves compatible matches, and places them on a multi-track timeline; as material accumulates, Reflector analyzes co-sounding regions, embeds constituent samples in a shared harmonic space, and aggregates them into session-level representations that update as material is added, removed, or transposed. The composer is never bound to the initial query as an anchor; any audio can be placed on the timeline, the query sample can be updated freely, and the session representation adapts around what is actually present.

This paper proceeds as a systems account. Section~\ref{sec:oracle} presents the compatibility oracle. Section~\ref{sec:training} describes the synthetic training pipeline that distills the oracle into an encoder. Section~\ref{sec:process} follows the system through indexing, session analysis, and cross-session navigation. Section~\ref{sec:characterization} characterizes the system's behavior on a working library of 631 samples: the fidelity of the embedding to its oracle, the location of its retrieval surface among classical tonal-similarity measures, and how retrieval follows arrangements through harmonic pivots and across whole sessions.

\section{The Compatibility Oracle}\label{sec:oracle}

At the center of Reflector is a hand-designed $12\times12$ table of interval weights, $K$. For any two sounds, $K$ scores how their pitch-class content interacts when layered. The score is computed frame by frame across time. This rule is the oracle. Every layer of the architecture either learns from it, aggregates with it, or approximates it.

\paragraph{Frame-level scoring.} We adapt the interval-class summation framework of Hall et al.~\cite{Hall2025} from intra-chord consonance to cross-object compatibility. Given two chroma frames $\mathbf{c}, \mathbf{d} \in \mathbb{R}^{12}$, the score is a kernel-weighted bilinear form:
\begin{equation}\label{eq:frame}
s_{\text{frame}} = \frac{\mathbf{c}^\top K\,\mathbf{d}}{(\mathbf{1}^\top\mathbf{c})(\mathbf{1}^\top\mathbf{d}) + \varepsilon}
\end{equation}
where $K_{ab} = \omega(\mathrm{ic}(a,b))$ maps each pitch-class pair to the net weight of its interval class. The outer product $\mathbf{c}\mathbf{d}^\top$ enumerates every pitch-class pair in which both sources are active; $K$ weights each pair by the interval class between them; the sum of the weighted couplings is the bilinear form $\mathbf{c}^\top K\,\mathbf{d}$. Where Hall et al.~\cite{Hall2025} score intervals within a single chord, we score intervals between two independent sources: internal consonance or dissonance within either source is invisible to the score, which responds only to the relationship the two objects create together. Normalizing by the product of total activation masses makes the score invariant to level while preserving relative activation balance, so the score reads as the average interval quality of the combination rather than its loudness.

\paragraph{Kernel weights.}
\begin{table}[h]
\centering
\small
\begin{tabular}{lcccr}
\toprule
IC & Interval & Cons. & Tens. & Net \\
\midrule
0 & P1/P8 & 1.0 & --- & $+1.0$ \\
1 & m2/M7 & --- & 1.0 & $-1.0$ \\
2 & M2/m7 & --- & 0.3 & $-0.3$ \\
3 & m3/M6 & 0.7 & --- & $+0.7$ \\
4 & M3/m6 & 0.7 & --- & $+0.7$ \\
5 & P4/P5 & 0.9 & 0.3 & $+0.6$ \\
6 & TT & --- & 0.8 & $-0.8$ \\
\bottomrule
\end{tabular}
\caption{Kernel weights used by Reflector.}
\label{tab:kernel}
\end{table}
The polarity of each interval class, and the magnitudes at classes 0, 1, and 5, follow the Schwartz-derived class-min baseline tabulated by Hall et al.~\cite{Hall2025}, in the interval-class tradition of Huron~\cite{Huron1994}; class 2 stays near that baseline. Thirds (classes 3 and 4) are elevated to encode shared tonality between co-sounding objects. Class 5 carries both consonance and tension weight, reflecting the contextual ambiguity of fourths and fifths between concurrent sources. The tritone is penalized more strongly than published intra-chord baselines as a marker of tonal incompatibility. The kernel is an experimental design position, stated in a table, and Reflector also exposes an unopinionated similarity lens (Section~\ref{sec:characterization}). We make no claim that these weights are perceptually optimal, or useful for every composer or library; Section~\ref{sec:characterization} measures what the choices do to the retrieval surface.

\paragraph{Trajectory-level scoring.} Sources are trajectories, so frame scores aggregate over time with interaction-mass weighting:
\begin{equation}\label{eq:traj}
s_{\text{traj}} = \sum_{t=1}^{T} w_t\, s_{\text{frame}}^{(t)}, \qquad w_t = \frac{\|\mathbf{c}_t\|_1 \, \|\mathbf{d}_t\|_1}{\sum_{u=1}^{T} \|\mathbf{c}_u\|_1 \, \|\mathbf{d}_u\|_1 + \varepsilon}
\end{equation}
Each moment contributes in proportion to how much both sources are actually sounding in it. A passage where the two objects overlap densely dominates the judgment; moments where either falls silent contribute nothing, so onsets, tails, and rests do not dilute the score. The frame weight and the frame score share the same interaction mass, one as numerator and one as denominator, and this shared structure admits an exact algebraic reduction of the whole trajectory score to a ratio of two sums; Section~\ref{sec:training} exploits it to make oracle scoring effectively free at runtime.

\paragraph{Transposition search.} The oracle evaluates the candidate under all twelve circular shifts of its pitch-class axis~\cite{Muller2007}:
\begin{equation}\label{eq:transposition}
k^* = \underset{k \in \{0,\ldots,11\}}{\arg\max}\; s_{\text{traj}}(\mathbf{C},\; \mathrm{roll}(\mathbf{D}, k))
\end{equation}
Rolling the candidate's chroma by $k$ semitones scores the combination as if the candidate were transposed; the argmax names the shift under which the pair blends best. The oracle emits the full twelve-score profile, and training supervises against the complete profile (Section~\ref{sec:training}), so the embedding learns whether a pair has one decisive alignment or several plausible ones. In Reflector, the profile surfaces directly: transposition suggestions name their interval, and the composer commits a direction by hand.

\section{Training Pipeline: Distilling the Oracle}\label{sec:training}

Because the system models pitch-class compatibility rather than multi-level musical properties, synthetic training data is a practical fit. We generate paired chroma trajectories from a pitch-class grammar and render them to audio through additive synthesis with domain randomization. The choice buys three things at once: full control over the training distribution, zero use of copyrighted recordings at any stage, and a pipeline that can be released as open source.

Training proceeds in two stages over different representations of the same labeled pairs. In both, the fundamental unit is a pair: a context and a candidate, each a $(T, 12)$ pitch-class activation trajectory at 50 fps with $T = 150$ frames (3 seconds). The oracle labels each pair with a compatibility score and a 12-way transposition profile.

\paragraph{Stage A} generates trajectories from pitch-class sets constructed over six triadic shapes (major, minor, diminished, augmented, sus2, sus4) with stochastic upper extensions and substitutions. Four temporal patterns (sustained chords, arpeggios, drone-plus-entry, layered polychords) provide structural variety, so the encoder encounters interval relationships in different temporal distributions. Activations are shaped by attack and release envelopes and compressed with a power-law exponent of 0.7, with no per-frame normalization, preserving the activation mass the oracle uses for frame weighting. A fraction of candidates receive stochastic chromatic ornaments at low activation strength, encoding the prior that transient chromatic activity should not dominate the compatibility signal. Two of the four temporal patterns use shared pitch-class sets, skewing labels toward high compatibility.

\paragraph{Stage B} renders each Stage A trajectory pair into audio via additive synthesis: waveform shape, detuned unison voices, reverb, chorus, distortion, and noise are randomized independently per source, with each pair assigned to a light or heavy effect-intensity tier. Chroma is re-extracted with a CQT (6 octaves, 36 bins per octave, $f_{\min} = 32.7$ Hz), pitch-class folding, and the same power-law compression. The CQT is a deliberate front-end choice: its log-spaced bins give constant resolution per semitone, so folding bins into pitch classes is exact across the full range, where a linear-frequency spectrogram smears low-octave semitones below bin resolution. Compression operates on raw synthetic activations before oracle scoring in Stage A and on CQT magnitudes in Stage B, so the oracle and the encoder see consistent dynamic range. Oracle labels carry forward from Stage A without recomputation: the encoder must recover the oracle's judgment from audio-derived chroma alone.

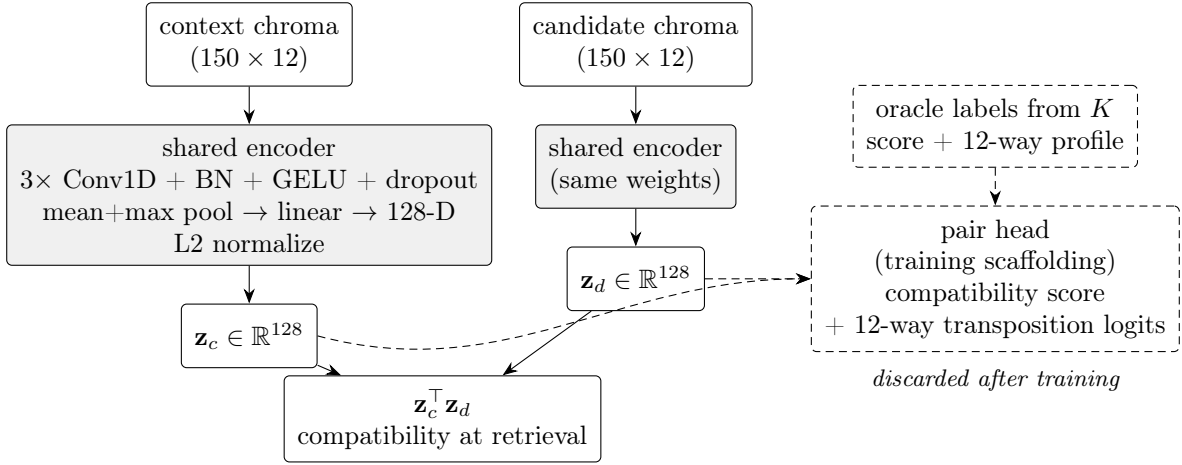
\begin{figure}[H]
\centering
\begin{tikzpicture}[
  font=\small,
  box/.style={draw, rounded corners=2pt, align=center, inner sep=5pt, minimum height=2.2em},
  enc/.style={box, fill=gray!12},
  scaf/.style={box, densely dashed},
  lbl/.style={font=\footnotesize\itshape},
  arr/.style={-{Stealth[length=2mm]}},
  node distance=5mm and 9mm
]
\node[box] (cin) {context chroma\\$(150 \times 12)$};
\node[box, right=22mm of cin] (din) {candidate chroma\\$(150 \times 12)$};

\node[enc, below=of cin] (ce) {shared encoder\\$3\times$ Conv1D + BN + GELU + dropout\\mean+max pool $\rightarrow$ linear $\rightarrow$ 128-D\\L2 normalize};
\node[enc, below=of din] (de) {shared encoder\\(same weights)};

\node[box, below=of ce] (zc) {$\mathbf{z}_c \in \mathbb{R}^{128}$};
\node[box, below=of de] (zd) {$\mathbf{z}_d \in \mathbb{R}^{128}$};

\node[box, below=9mm of $(zc)!0.5!(zd)$] (dot) {$\mathbf{z}_c^\top \mathbf{z}_d$\\compatibility at retrieval};

\node[scaf, right=14mm of zd, align=center] (head) {pair head\\(training scaffolding)\\compatibility score\\+ 12-way transposition logits};
\node[lbl, below=1mm of head] {discarded after training};

\node[scaf, above=of head] (oracle) {oracle labels from $K$\\score + 12-way profile};

\draw[arr] (cin) -- (ce);
\draw[arr] (din) -- (de);
\draw[arr] (ce) -- (zc);
\draw[arr] (de) -- (zd);
\draw[arr] (zc) -- (dot);
\draw[arr] (zd) -- (dot);
\draw[arr, densely dashed] (zc.east) to[out=-15, in=180] (head.west);
\draw[arr, densely dashed] (zd) -- (head);
\draw[arr, densely dashed] (oracle) -- (head);
\end{tikzpicture}
\caption{Two-tower architecture. Solid paths survive to inference; dashed paths are training scaffolding.}
\label{fig:model}
\end{figure}

\paragraph{Supervision.} Figure~\ref{fig:model} outlines the architecture. A shared encoder (three 1-D convolutions with batch normalization, GELU activations, and dropout; mean+max pooling; projection to 128-D; L2 normalization) maps context and candidate into one embedding space. A pair head receives the concatenation of both embeddings, their absolute difference, and their element-wise product, and produces a scalar compatibility prediction and 12-way transposition logits; the head is training scaffolding, discarded after Stage B. Compatibility is supervised by mean squared error against the oracle's best-over-transpositions score. Transposition is supervised by forward KL divergence against the oracle's full 12-way profile with temperature sharpening on both distributions:
\begin{equation}\label{eq:kl}
\mathcal{L}_{\text{KL}} = D_{\text{KL}}\Big(\text{softmax}(\mathbf{s} / \tau) \;\|\; \text{softmax}(\hat{\mathbf{l}} / \tau)\Big)
\end{equation}
where $\mathbf{s}$ is the oracle's score vector, $\hat{\mathbf{l}}$ are the model's transposition logits, and $\tau = 0.25$. The sub-unity temperature sharpens target and prediction alike, emphasizing the best transposition and penalizing diffuse outputs; forward KL penalizes under-coverage of oracle-preferred shifts. The Stage A loss sums the KL term and $\text{MSE}(\hat{c}, \max_k s_k)$ with equal weight, where $\hat{c}$ is the pair head's scalar output. Key-invariance augmentation (random circular shift, probability 0.5) discourages anchoring to absolute pitch-class identities. Stage A trains with AdamW ($\text{lr} = 1\times10^{-4}$, batch size 64, 8 epochs; 50k train / 5k validation / 5k test pairs) on an Apple M4.

\paragraph{Retrieval fine-tuning.} Stage B initializes from the Stage A checkpoint, resets the optimizer, and switches to InfoNCE~\cite{vandenOord2018} on audio-derived chroma, directly optimizing the dot-product geometry used at inference:
\begin{equation}\label{eq:infoNCE}
\mathcal{L}_{\text{NCE}} = -\frac{1}{N}\sum_{i=1}^{N}\log \frac{\exp(\mathbf{z}_{c_i}^\top \mathbf{z}_{d_i} / \tau_r)}{\sum_{j=1}^{N}\exp(\mathbf{z}_{c_i}^\top \mathbf{z}_{d_j} / \tau_r)}
\end{equation}
with $\tau_r = 0.2$. A small MSE anchor ($\lambda = 0.05$) on the pair head's scalar output prevents the embedding's scores from drifting away from the oracle's judgments. The tether holds the embedding's values. Key-rotation augmentation is not applied; the encoder inherits key invariance from Stage A. The encoder is frozen after this stage, permanently: every representation in Reflector is a coordinate in one fixed space.

\paragraph{Scoring at scale: an exact reduction.} The oracle scores sixty thousand trajectory pairs during data generation, and indexing asks it to score every pair of windows within every indexed file. Computed naively, Equations~\ref{eq:frame} and~\ref{eq:traj} cost a 144-cell weighted sum per frame pair; a five-minute file's window set alone implies roughly $4\times10^8$ interpreted operations. By substituting Equation~\ref{eq:frame} into Equation~\ref{eq:traj}, each frame's score is divided by the same interaction mass that Equation~\ref{eq:traj} multiplies it by, and the normalizations cancel term by term:
\begin{equation}\label{eq:reduction}
s_{\text{traj}} = \frac{\sum_{t=1}^{T} \mathbf{c}_t^\top K\,\mathbf{d}_t}{\sum_{u=1}^{T} (\mathbf{1}^\top\mathbf{c}_u)(\mathbf{1}^\top\mathbf{d}_u)}
\end{equation}
The trajectory score is one ratio of two bilinear accumulations. Stacking the $W$ windows of a file as a tensor $X \in \mathbb{R}^{W \times T \times 12}$, flattening both $XK$ and $X$ to $W \times 12T$ matrices makes every pairwise numerator a single matrix product, and the denominators are the outer product of per-window activation masses: two matrix multiplications for all $W(W-1)/2$ pairs across all frames. In the shipped implementation it evaluates a 198-window file's complete pairwise coherence in about two milliseconds where the direct loop takes about a minute, a speedup of four orders of magnitude. 

\section{System Process}\label{sec:process}

\subsection{Indexing and Retrieval}\label{subsec:indexing}

To index a sample library, the system extracts full-length CQT chroma from each file, segments it into overlapping 3-second windows matching the training distribution (50\% overlap), and embeds each window with the frozen encoder. The pre-normalization embedding magnitude serves as a confidence signal: near-zero projections are discarded, since L2 normalization would inflate them to unit length without meaningful content. The surviving windows are weighted by mean pairwise oracle coherence, computed through Equation~\ref{eq:reduction} at negligible cost, so harmonically coherent windows contribute more to the file representation. The file embedding is the L2-normalized weighted mean. The aggregated chroma trajectory is stored alongside the embedding for downstream oracle scoring during session analysis, and a precomputed dot-product matrix across all indexed files makes within-library retrieval a single array lookup.

Retrieval operates in two phases. Given a query, the system ranks all indexed candidates by dot product and returns the top-10 as direct compatibility matches. A second, optional phase evaluates the remaining candidates under all twelve pitch-class transpositions: raw chroma is circularly shifted, re-embedded, and scored against the query embedding. On the working library, 68\% of top-10 query pairs achieve higher compatibility under a non-zero transposition, which is the phase's reason to exist: a sample that ranks poorly in its original key may be a strong match under pitch shift. Transposition results arrive progressively, so the composer can act on early matches without waiting for the full sweep.

\subsection{Session Structure}\label{subsec:session}

\begin{figure}[H]
  \centering
  \includegraphics[width=\linewidth]{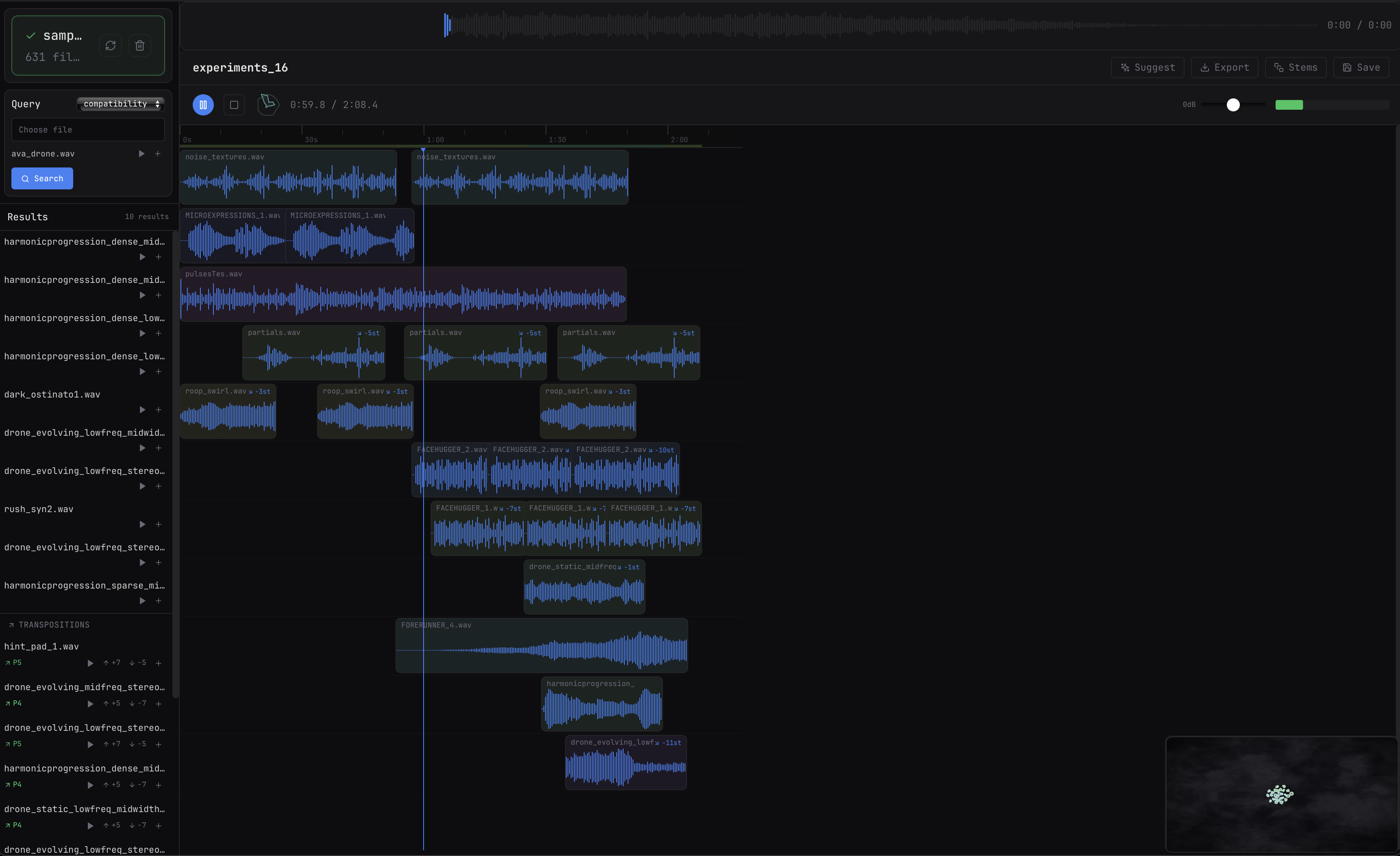}
  \caption{Session view: a multi-track arrangement with query results, transposition matches, and per-placement transposition shifts.}
  \label{fig:session-view}
\end{figure}

The composer arranges samples on a multi-track timeline (Figure~\ref{fig:session-view}). Reflector maintains two representations of the session's harmonic identity: a live arrangement centroid, recomputed from the current placements whenever the composer requests suggestions, and a saved session centroid.

On each save, a sweep-line pass identifies every distinct time region where the set of co-sounding samples changes. Within each region, the active members' embeddings combine into a region centroid weighted by their mean pairwise oracle coherence. Region centroids then aggregate by duration into the session centroid, so passages that persist longer exert greater influence on the session's harmonic identity; the result is L2-normalized to the unit hypersphere. Each centroid carries a dispersion value, the coherence-weighted variance of its members' embeddings around it, measuring how tightly the material clusters in the embedding space. This saved centroid is the authoritative session representation: it is stored in the database, projected into 3-D for the galaxy view (Section~\ref{subsec:galaxy}), and used for nearest-neighbor session similarity. 

\subsection{Galaxy}\label{subsec:galaxy}

\begin{figure}[H]
  \centering
  \includegraphics[width=\linewidth]{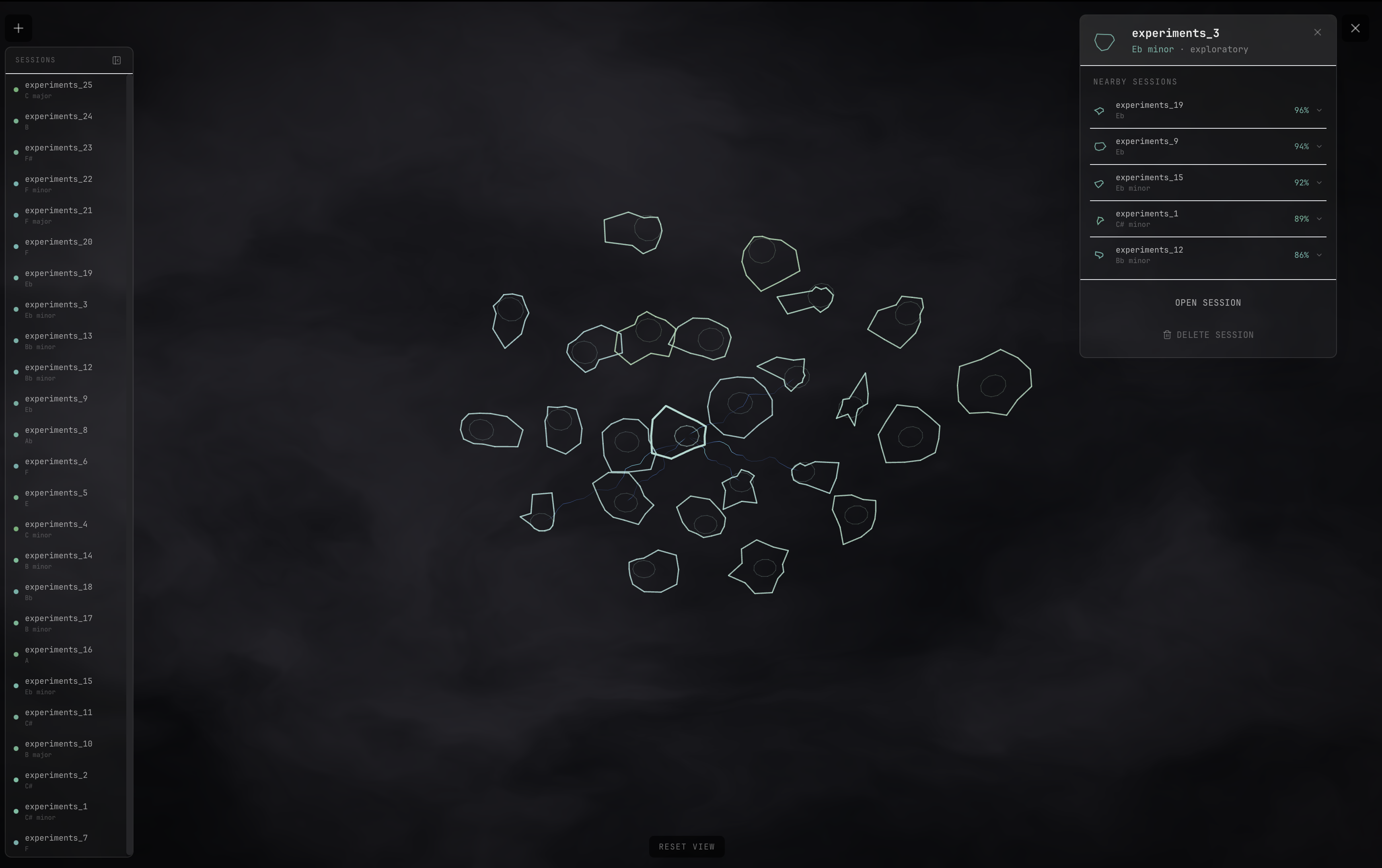}
  \caption{Galaxy view: session centroids in a navigable 3-D space, with nearest neighbors ranked by cosine similarity.}
  \label{fig:galaxy-view}
\end{figure}

Session centroids are projected from 128-D to 3-D for layout stability (Figure~\ref{fig:galaxy-view}). Each node represents a session; nearest neighbors surface which sessions share harmonic vocabulary. Selecting a node displays its tonal center, dispersion category, and inter-session compatibility scores. The composer can act on these connections directly, incorporating samples from one session into another, as the taxonomy reveals relationships between existing work that no single query would surface.

\section{System Characterization}\label{sec:characterization}

This section describes how the system behaves; it does not argue that the behavior is superior to existing or potential alternatives. The measurements are taken on the working corpus the system was developed against: 631 samples drawn from the author's and another professional composer's personal collections spanning electronic, acoustic, field recording, textural, and processed synthesis, with durations from seconds to minutes. Sample libraries are curated by nature, and every result below is a property of this corpus as seen through these rules; we make no generality claim beyond it.

\subsection{Retrieval Surface Analysis}\label{subsec:surface}
\begin{figure}[H]
  \centering
  \includegraphics[width=\linewidth]{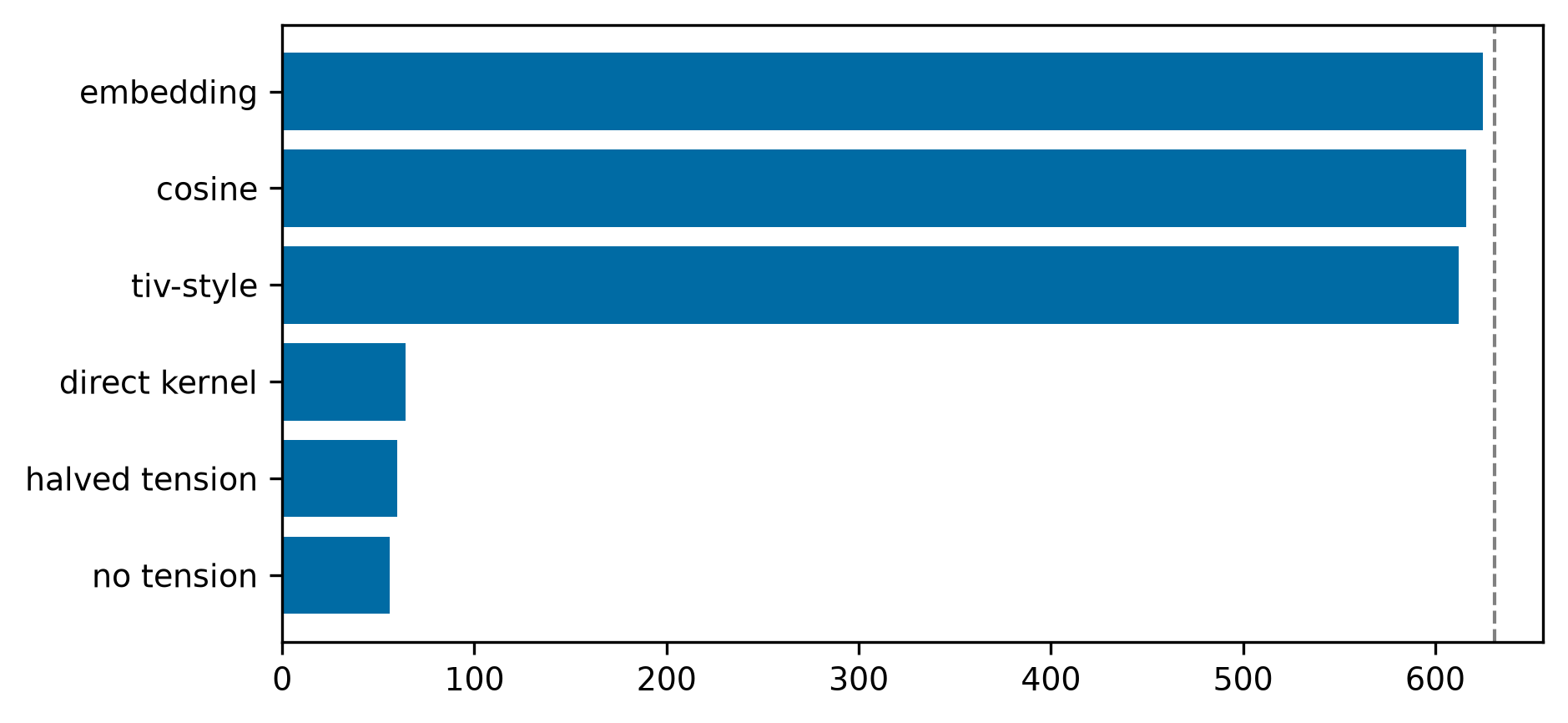}
  \caption{Retrieval surface coverage for all six rules: how many of the 631 samples each ever surfaces in a top-10.}
  \label{fig:coverage}
\end{figure}

\paragraph{Protocol.} To locate the kernel among established alternatives, we rank the full corpus under six rules and compare which samples appear in each rule's top-10, order ignored, over all 631 queries. The embedding is the learned surface Reflector ships. Cosine is mean-chroma cosine, the classical chroma-similarity measure. TIV-style is Euclidean distance over DFT (Discrete Fourier Transform) coefficients 1--6 of mean chroma. The direct kernel is the kernel of Table~\ref{tab:kernel} scored on stored chroma trajectories, best over 12 transpositions, with the encoder bypassed; halved tension and no tension are the same scoring with the kernel's tension weights halved and removed.

\paragraph{Coverage collapse under direct kernel scoring.} Ranked by the direct kernel, only 64 of 631 samples ever appear in any top-10 (Figure~\ref{fig:coverage}), and it was found that a single sparse consonant drone sample appears in 97\% of them. A quiet, tonally pure drone blends acceptably with nearly everything, and with twelve transpositions available to align itself, it wins almost every contest. The rule judges each pair in isolation and has no concept of variety, so it crowns universal donors. The perturbed kernels concentrate identically (halved tension covers 60 samples, no tension 56), so the degeneracy belongs to interval-weighted direct scoring, and softer weights do not soften it. The direct kernel is not a retrieval path in the application; we score it only to measure what follows.

\paragraph{Coverage recovery in the learned embedding.} The embedding preserves the kernel's values while abandoning its selections. With each indexed sample as query, NDCG (normalized discounted cumulative gain) against the oracle's best-over-transpositions score is 0.85 at rank 10 (standard error 0.003 across queries; no query below 0.73), well beyond the encoder's synthetic training domain; yet the embedding's top-10s share only 1.3\% of their members with the direct kernel's, and its surface covers 625 of 631 samples with a maximum residency of 29 (Figure~\ref{fig:coverage}). We attribute the escape to two properties of the architecture. Scores must be expressed as dot products of L2-normalized vectors, and on the unit sphere no candidate can be close to every query direction at once: a candidate's mean score across the library, $\bar{\mathbf{z}}^{\top}\mathbf{z}_d$, is bounded by $\|\bar{\mathbf{z}}\|$, the norm of the mean query embedding, which is small whenever the library is spread over the sphere. A universal donor is therefore inexpressible: normalization deletes the degree of freedom the degeneracy requires. InfoNCE reinforces this, since a candidate scoring high against every in-batch context inflates the denominator everywhere and is pushed away from contexts it is not paired with. This pressure is the uniformity property of contrastive learning: InfoNCE asymptotically optimizes alignment of positive pairs jointly with uniformity of the embedding distribution on the hypersphere~\cite{WangIsola2020}, and uniform spread is the configuration in which no single direction scores well against the whole collection. The approximation error therefore lands on the degenerate component of the rule, and what survives is the interaction structure.

\begin{figure}[H]
  \centering
  \includegraphics[width=\linewidth]{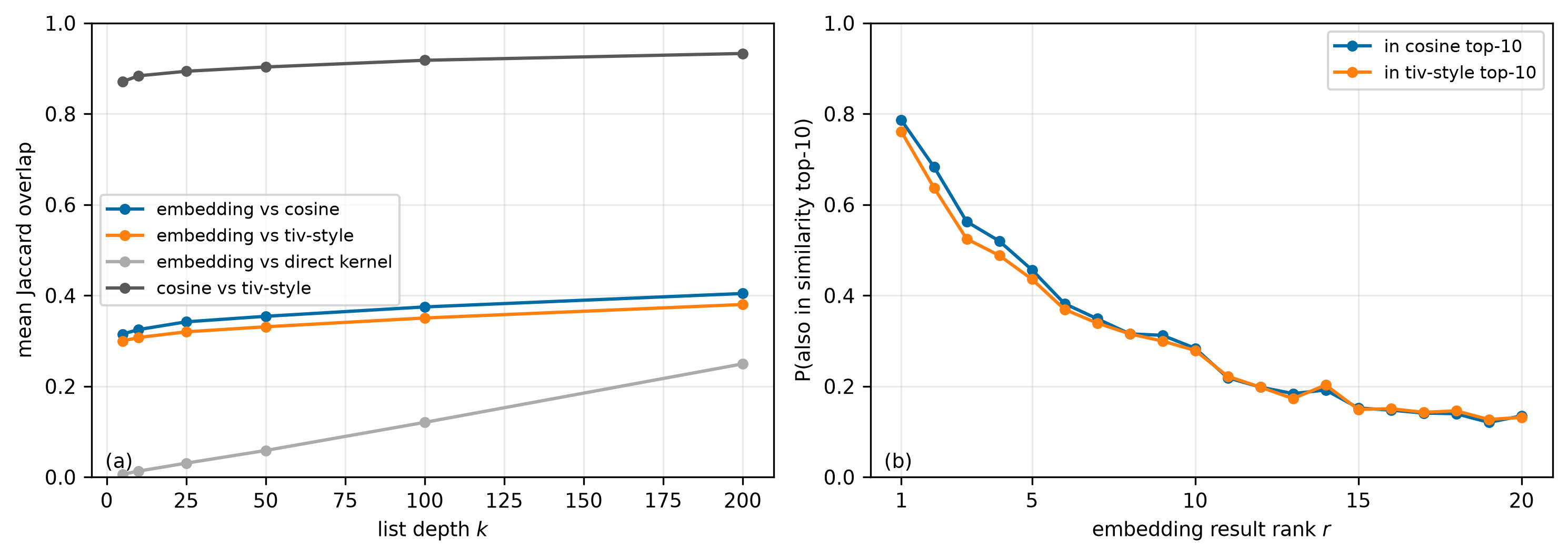}
  \caption{(a) Mean top-$k$ Jaccard agreement for the principal rule pairs as a function of list depth, over all 631 queries. (b) Probability that the embedding's rank-$r$ result appears in the similarity measures' top-10.}
  \label{fig:surface}
\end{figure}

\paragraph{Similarity and compatibility families.} Similarity measures answer what resembles the query: they retrieve its siblings, versions, and near relatives, and they behave as one family. Cosine and TIV-style overlap $0.88 \pm 0.12$ (standard deviation across queries; Figure~\ref{fig:surface}a), an agreement that is partly guaranteed by construction: coefficients 1--6 of a 12-point real signal omit only the DC term, so by Parseval's theorem the unweighted DFT distance nearly reproduces chroma distance, and the published TIV departs from cosine chiefly through its perceptual coefficient weights. The kernel answers what combines with the query, and it retrieves complements. On this corpus the two surfaces share 0.4\% of their top-10 members and their full rankings are anti-correlated (mean Spearman $\rho = -0.65$), with the kernel's favorites sitting near the bottom of cosine's list. These are answers to two different questions.

\paragraph{Consensus decay.} The consensus-decay curve (Figure~\ref{fig:surface}b) locates the embedding relative to the similarity family: its first-ranked sample appears in the similarity measures' top-10 for 79\% of queries, its tenth for 28\%, so the top ranks hold the matches every rule finds and the lower ranks hold the kernel's preference.

\paragraph{Summary.} Three findings. Direct scoring with any kernel in the family collapses coverage onto a few universal donors, and adjusting the weights does not repair it. Similarity measures cover the library but rank it nearly opposite to the kernel: resembling and combining are different questions. The learned embedding is the only surface that carries the kernel's values and covers the library at once, because its normalized geometry cannot express the degenerate component of the rule. The design consequence follows.

\paragraph{Two surfaces in the application.} Because the surfaces are near-disjoint, exposing both gives each query two nearly non-overlapping result lists from the same corpus at zero training cost. Reflector offers two lenses for querying a sample: the oracle-trained embedding, answering what combines well with the query; and cosine over the index's stored mean chroma, answering what resembles it, for locating versions, alternate takes, and family. The similarity lens is the unopinionated identity-kernel special case of the same bilinear form; the default lens carries the designed preference. Session analysis, centroids, and the suggestion engine always use the embedding.

\subsection{Arrangement-Aware Analysis}\label{subsec:arrangement}

The analysis above treats retrieval from a single fixed query. Reflector's defining behavior is that retrieval tracks the arrangement as it evolves: each sample placed, removed, or transposed shifts the session representation, and the suggestions shift with it. This section examines that behavior in two views, a deliberate harmonic pivot within one session and the accumulated drift across many sessions.

\begin{figure}[H]
  \centering
  \includegraphics[width=\linewidth]{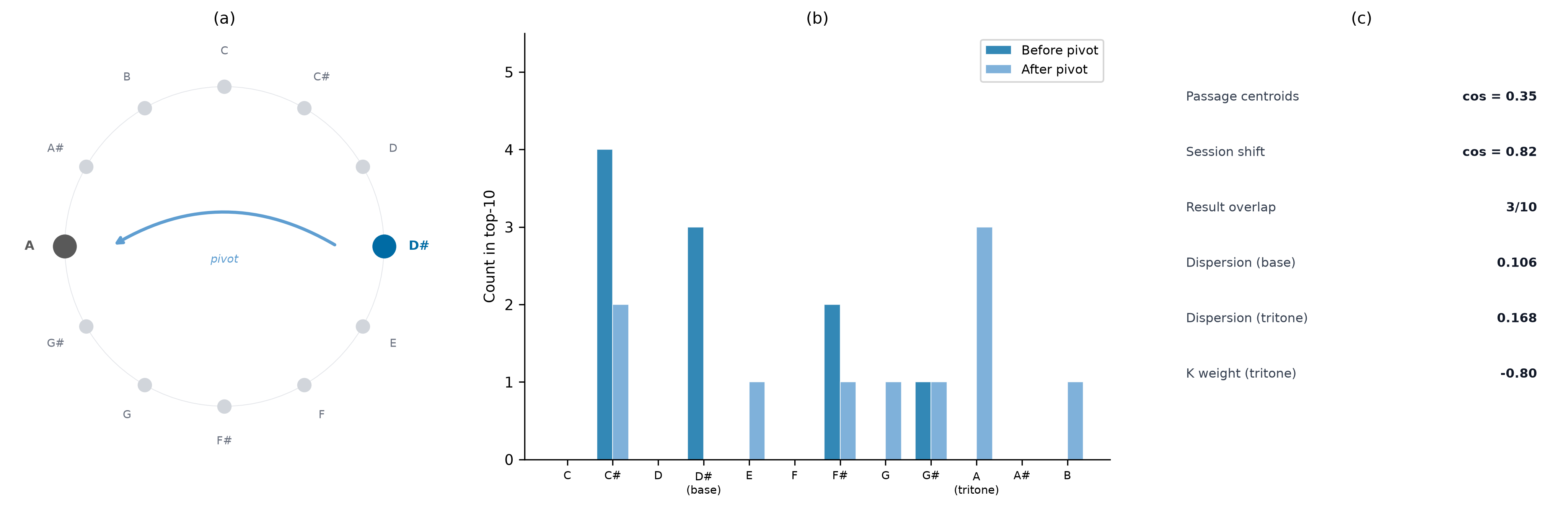}
  \caption{Retrieval behavior before and after a tritone pivot.}
  \label{fig:tritone}
\end{figure}

\paragraph{Tritone pivot.} A session begins with material centered on D$\sharp$ and pivots into co-sounding regions an A--D$\sharp$ tritone apart (Figure~\ref{fig:tritone}). The tonal center migrates on the pitch-class circle and retrieval redistributes with it: 70\% of top-10 suggestions change, results centered on the base key leave the list entirely, and samples centered on A enter it. The two passage centroids occupy distinct regions of the embedding space ($\cos = 0.35$), and the session centroid moves partway between them ($\cos = 0.82$ against its pre-pivot position), the duration-weighted blend the sweep-line produces. Dispersion rises from 0.106 in the consonant base to 0.168 in the tritone passage: the library holds fewer samples that cohere in that region, so the material forms a broader neighborhood. The system follows the composer into territory its own rule penalizes, which is the intended division of labor.

\begin{figure}[H]
  \centering
  \includegraphics[width=\linewidth]{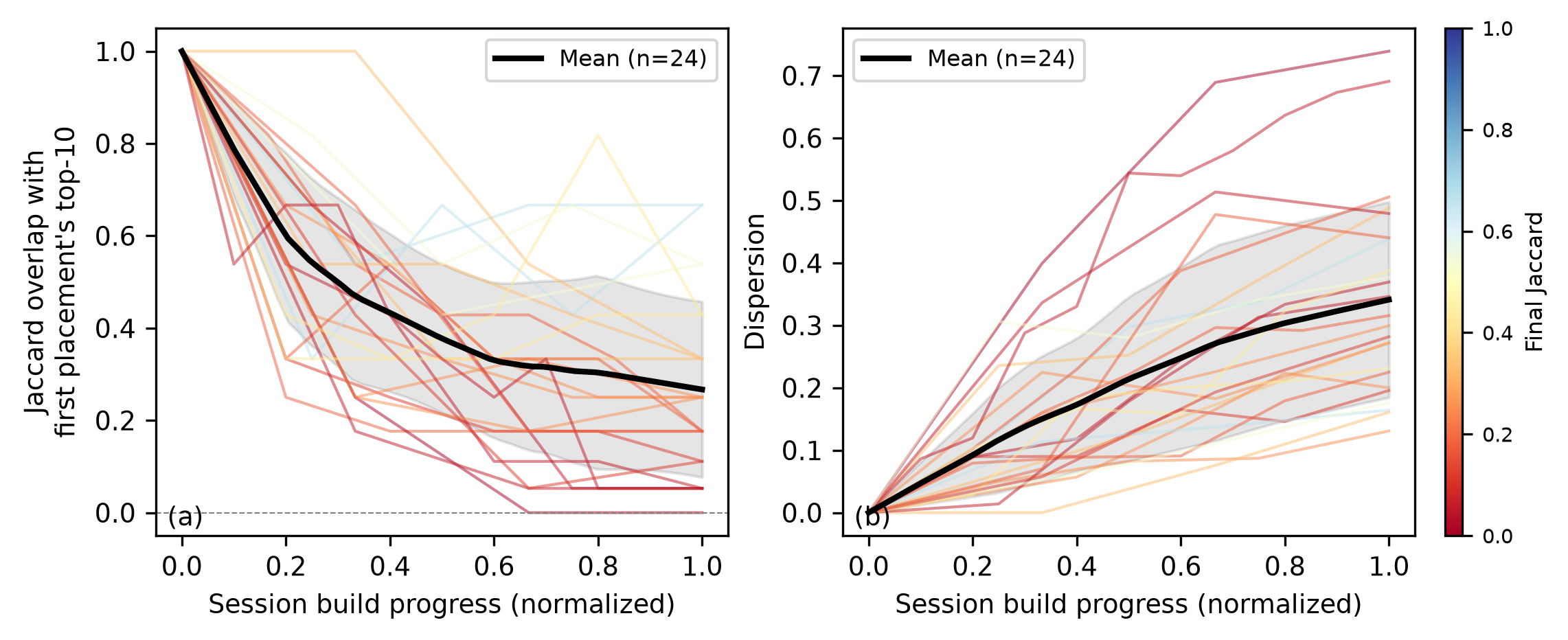}
  \caption{Centroid drift across 24 composer sessions. (a) Top-10 suggestion overlap between the evolving arrangement centroid and the session's first placed sample. (b) Dispersion as material accumulates. Trajectories colored by final overlap.}
  \label{fig:drift}
\end{figure}

\paragraph{Centroid drift.} Across 24 sessions from the same two composers, one of whom is the author (Figure~\ref{fig:drift}), placements are replayed in timeline order and the evolving arrangement centroid's top-10 suggestions are compared against the top-10 of the session's first placed sample. Mean overlap declines from 1 to 0.27 as sessions develop: roughly three quarters of suggestions change over the course of an arrangement, with the steepest divergence in the first fifth of session progress, where each placement moves the centroid most (mean overlap is already 0.63 at one fifth). Dispersion rises monotonically as material accumulates in every session analyzed, and individual trajectories vary widely, with final overlaps ranging from 0 to 0.67. These curves show that retrieval anchored to a session rather than a single sample has something to say precisely because the two diverge musically.

\section{Limitations}\label{sec:limitations}

Reflector is built on Western tonal harmony. The kernel weights and the synthetic training grammar encode its consonance conventions, with thirds deliberately elevated; adapting the system to other intervallic traditions requires a different weight table and dataset design. The architecture permits this, since $K$ is a parameter, but the work is not explored here.

The embedding captures pitch-class relationships only. A harmonically compatible sample may be unsuitable in timbre, texture, dynamics, or rhythm; those judgments remain with the composer. The similarity lens shares a related coarseness from the other direction: cosine over mean chroma collapses time entirely, which suits its family-finding purpose and nothing subtler.

The kernel's departures from published interval tables remain unvalidated against listening data. The surface analysis of Section~\ref{sec:characterization} locates the system's taste but does not justify it. A perceptual study, with the tonal-similarity descriptors of Section~\ref{sec:introduction} as reference points, is the appropriate instrument for that question and is left open: Reflector is currently offered as an application whose validation is use.

Section~\ref{sec:characterization} exposed a boundary: the direct kernel's retrieval surface is degenerate, concentrating on universal-donor material, though the shipped system escapes it. The mitigation emerges from normalization and the contrastive objective, choices made against overfitting to the oracle during development; its role as diversity control was not measured until now. Explicit diversity control in retrieval is unexplored.

All characterization is on a single 631-sample corpus. Each indexed file is represented by one embedding aggregated across windows and dominated by its most harmonically stable content; atonal material or music that changes key frequently may be misrepresented, and chunk-level retrieval, returning temporal regions rather than whole files, may help address this. Files must be indexed to participate in session analysis. The training set is fixed at 50k pairs, and the effect of scale is unexplored.

\section{Conclusion}\label{sec:conclusion}

Reflector demonstrates that a fixed harmonic oracle, grounded in interval-class theory, with an encoder trained on synthetic audio alone, can support arrangement-aware retrieval that responds to compositional decisions in real time. A single embedding space supports retrieval at every level of the compositional process: sample-to-sample, arrangement-to-library, and session-to-session. The system provides a stable coordinate system through which the composer's evolving harmonic practice becomes visible. Every suggestion is a deterministic function of what the composer has placed on the timeline, flowing through a geometry they can interrogate at every level: from pairwise compatibility scores to session centroids to cross-session topology.

The characterization taught the design something it did not know. Scored directly as a retrieval rule, a configuration the shipped system never runs, the kernel concentrates its top ranks onto a handful of universal donors; the learned embedding covers the library because its normalized geometry cannot express the degeneracy. Distillation acted as a filter: the kernel's interaction structure survived compression into the embedding, while the degenerate component, which the unit sphere cannot carry, was left behind. In addition, similarity and compatibility proved to be nearly opposite orderings of the same collection, a distinction Reflector exposes as two lenses on one library.

A curated sample library contains latent compositional structure: relationships between files that no manual workflow can fully enumerate and no single query can reveal. Reflector makes that structure navigable and tracks how the composer moves through it.

\section*{Ethics Statement}
Reflector was designed with a deliberate ethical commitment. The contrastive encoder was trained entirely on synthetic audio; no copyrighted or commercially licensed material was used at any stage of development. This decision reflects the legal and ethical uncertainty surrounding music AI systems trained on existing recordings.

\section*{Availability}
Reflector is available as a free, signed macOS application at \url{https://github.com/austinrockman/reflector}. The training and preprocessing pipeline is open source at the same link.


\begin{thebibliography}{99}

\bibitem{Chen2020}
B.-Y.~Chen, J.~B.~L.~Smith, and Y.-H.~Yang, ``Neural loop combiner: Neural network models for assessing the compatibility of loops,''
in {\em Proc. Int. Society for Music Information Retrieval Conf.}, pp.~424--431, 2020.

\bibitem{Lattner2022}
S.~Lattner, ``SampleMatch: Drum sample retrieval by musical context,''
in {\em Proc. Int. Society for Music Information Retrieval Conf.}, pp.~781--788, 2022.

\bibitem{Gebhardt2016}
R.~B.~Gebhardt, M.~E.~P.~Davies, and B.~U.~Seeber, ``Psychoacoustic approaches for harmonic music mixing,''
{\em Applied Sciences}, vol.~6, no.~5, p.~123, 2016.

\bibitem{Riou2024}
A.~Riou, S.~Lattner, G.~Hadjeres, M.~Anslow, and G.~Peeters, ``Stem-JEPA: A joint-embedding predictive architecture for musical stem compatibility estimation,''
in {\em Proc. Int. Society for Music Information Retrieval Conf.}, pp.~625--633, 2024.

\bibitem{Sononym}
Sononym, ``About Sononym,'' 2024. [Online]. Available: https://www.sononym.net/about/sononym/

\bibitem{XLNXO}
XLN Audio, ``XO,'' 2024. [Online]. Available: https://www.xlnaudio.com/products/xo

\bibitem{SpliceCreate}
Splice, ``The future of music creation is here,'' 2024. [Online]. Available: https://splice.com/innovation/timeline

\bibitem{OutputCoProd}
Output, ``Co-Producer User Guide,'' 2025. [Online]. Available: https://support.output.com/en/articles/10628997-co-producer-user-guide

\bibitem{Gomez2006}
E.~G\'omez, ``Tonal description of polyphonic audio for music content processing,''
{\em INFORMS Journal on Computing}, vol.~18, no.~3, pp.~294--304, 2006.

\bibitem{Harte2006}
C.~Harte, M.~Sandler, and M.~Gasser, ``Detecting harmonic change in musical audio,''
in {\em Proc. 1st ACM Workshop on Audio and Music Computing Multimedia}, pp.~21--26, 2006.

\bibitem{Bernardes2016}
G.~Bernardes, D.~Cocharro, M.~Caetano, C.~Guedes, and M.~E.~P.~Davies, ``A multi-level tonal interval space for modelling pitch relatedness and musical consonance,''
{\em Journal of New Music Research}, vol.~45, no.~4, pp.~281--294, 2016.

\bibitem{Muller2007}
M.~M\"uller and M.~Clausen, ``Transposition-invariant self-similarity matrices,''
in {\em Proc. Int. Society for Music Information Retrieval Conf.}, pp.~47--50, 2007.

\bibitem{Brown1991}
J.~C.~Brown, ``Calculation of a constant Q spectral transform,''
{\em Journal of the Acoustical Society of America}, vol.~89, no.~1, pp.~425--434, 1991.

\bibitem{BrownPuckette1992}
J.~C.~Brown and M.~S.~Puckette, ``An efficient algorithm for the calculation of a constant Q transform,''
{\em Journal of the Acoustical Society of America}, vol.~92, no.~5, pp.~2698--2701, 1992.

\bibitem{Huron1994}
D.~Huron, ``Interval-class content in equally tempered pitch-class sets: Common scales exhibit optimum tonal consonance,''
{\em Music Perception}, vol.~11, no.~3, pp.~289--305, 1994.

\bibitem{Hall2025}
E.~T.~R.~Hall, R.~Tamir, and M.~Rohrmeier, ``Optimising computational measures from behavioural data predicts perceived consonance,''
{\em Trans. Int. Society for Music Information Retrieval}, vol.~8, no.~1, pp.~313--333, 2025.

\bibitem{vandenOord2018}
A.~van~den~Oord, Y.~Li, and O.~Vinyals, ``Representation learning with contrastive predictive coding,''
{\em arXiv preprint arXiv:1807.03748}, 2018.

\bibitem{WangIsola2020}
T.~Wang and P.~Isola, ``Understanding contrastive representation learning through alignment and uniformity on the hypersphere,''
in {\em Proc. Int. Conf. on Machine Learning}, pp.~9929--9939, 2020.

\end{thebibliography}
\end{document}